\documentclass[12pt]{article}
\usepackage{graphicx}
\renewcommand{\bar}[1]{\overline{#1}}
\renewcommand{\L}{{\cal L}}
\newcommand{\M}{{\cal M}}
\newcommand{\V}{{\cal V}}

\newcommand{\lsim} {\buildrel < \over {_\sim}}
\newcommand{\gsim} {\buildrel > \over {_\sim}}
\def\ie{{\it i.e.}}
\def\hc{{\rm h.c.}}
\def\eg{{\it e.g.}}
\renewcommand{\thefootnote}{\fnsymbol{footnote}}
\begin{document}

\setcounter{page}{1} \thispagestyle{empty}
\rightline{\vbox{\halign{&#\hfil\cr
                         &SLAC-PUB-9212\cr
                         &UH-511-1000-02\cr
                         &May 2002\cr}}}
\vspace{0.8in}
\begin{center}

{\Large\bf Yet Another Extension of the Standard Model:  Oases in
the Desert?} \footnote{Work supported by the Department of Energy,
contract DE-AC03-76SF00515 and grant DE-FG-03-94ER40833.}
\\ \bigskip
{\normalsize \large J. D. Bjorken, S. Pakvasa and S. F. Tuan}
\\ \smallskip
{\it {Stanford Linear Accelerator Center\\
Stanford University, Stanford, CA 94309}}\\
and
{\it {University of Hawaii, Honolulu, HI 96822}}\\
\end{center}

\vfill
\begin{center}

Submitted to Physical Review {\bf D}.

\end{center}
\vfill
\newpage

\begin{center} ABSTRACT
\end{center}

We have searched for conceptually simple extensions of the
standard model, and describe here a candidate model which we find
attractive.  Our starting point is the assumption that
off-diagonal CKM mixing matrix elements are directly related by
lowest order perturbation theory to the quark mass matrices. This
appears to be most easily and naturally implemented by assuming
that all off-diagonal elements reside in the down-quark mass
matrix.  This assumption is in turn naturally realized by
introducing three generations of heavy, electroweak-singlet down
quarks which couple to the Higgs sector diagonally in flavor,
while mass-mixing off-diagonally with the light down-quarks.
Anomaly cancellation then naturally leads to inclusion of
electroweak vector-doublet leptons.  It is then only a short step
to completing the extension to three generations of fundamental
$\underline{27}$ representations of $E(6)$.

Consequences of this picture include (1) the hypothesis of ``Stech
texture'' for the down-quark mass matrix (imaginary off-diagonal
elements) leads to an approximate right unitarity triangle
($\gamma\approx \pi/2$), and a value of $\sin 2\beta$ between 0.64
and 0.80; (2) Assuming only that the third generation $B$ couples
to the Higgs sector at least as strongly as does the top quark,
the mass of the $B$ is roughly estimated to lie between 1.7 TeV
and 10 TeV, with lower-generation quarks no heavier.  The
corresponding guess for the new leptons is a factor two lower, 0.8
TeV to 5 TeV; (3) Within the validity of the model, flavor and CP
violation are ``infrared'' in nature, induced by semi-soft mass
mixing terms, {\em not} Yukawa couplings; (4)  The ``Mexican hat''
structure of the Higgs potential may be radiatively induced by the
new heavy down-quark one-loop contributions to the potential; (5)
A subset of the precision electroweak experiments are sensitive to
the physics induced by the heavy quarks and/or leptons; (6) If the
Higgs couplings of the new quarks are flavor symmetric, then there
necessarily must be at least one ``oasis'' in the desert, induced
by new radiative corrections to the top quark and Higgs coupling
constants, and roughly at  $10^3$ TeV; (7) If the Higgs couplings
of the new down-quarks are hierarchical and equal to the usual
Higgs couplings to right-handed up quarks, then, in the limit in
which gauge coupling constants are set to zero,  parity violation
is also ``infrared", induced by semi-soft mass terms.

\newpage

\renewcommand{\thefootnote}{\arabic{footnote}}

\section{Introduction}

The problem of extending the standard model \cite{ref:1} needs no
motivation. Attempted extensions are legion, and most are
complicated. Many nowadays are based on the hypothesis of low
energy supersymmetry \cite{ref:2}, where a wealth of new
superpartners, each with an uncertain phenomenology, and more than
one hundred new parameters are introduced. Attempts to do without
supersymmetry generically end up with a proliferation of new Higgs
representations, a variety of new fermions and gauge bosons with
uncertain phenomenology \cite{ref:3}, and less of an underlying
esthetic than possessed by the supersymmetric models.

The minimal standard model, with its single Higgs boson as the
only undiscovered element, remains at present the unique
description which is both very simple and highly credible.  The
``desert"  scenario of no new physics between the electroweak and
grand unified theory (GUT) scales has a completely defined Hilbert
space and a consistent, calculable S-matrix at all energy scales
up to the GUT scale \cite{ref:4}, provided the Higgs mass  lies in
the relatively narrow window of $160 \pm 20$ GeV. And it is
certainly arguable that the hierarchy problem, \ie\ the fate of
the quadratically divergent renormalization of the Higgs boson
mass \cite{ref:5}, is so similar to the cosmological-constant
problem \cite{ref:6} that the best thing to do is to treat it in
the same way, namely to set it aside as ``not understood",
assuming it to be solved later at a much deeper level. But even
within the minimal standard model, the perturbative theory will
break down at a mass scale less than the GUT scale unless the
Higgs mass is within the aforementioned window. If something like
this turns out to be the scenario, then there will be at least one
``oasis" in the desert, a landmark mass scale where new physics
and most likely new strong forces emerge \cite{ref:7}.

It is reasonable in fact to define the problem of extending the
standard model by this question: ``What is the mass scale of the
first oasis, and what is its particle content?"  The minimal
supersymmetric standard model defines that mass scale as
the electroweak scale itself, with subsequent oases far away and
poorly defined (hidden and/or gauge sectors). Technicolor models
\cite{ref:8} put oases at the TeV scale, with others higher up
(extended technicolor), although all are plagued with
phenomenological constraints difficult to satisfy \cite{ref:9}.

\eject

 In attacking the problem again, we restrict ourselves to
conceptually simple, reasonably well-motivated extensions.  We try
not to force the model into preconceived ideology, but instead let
the structure of the model lead to the next step until a dead end
occurs, where no simple extension can be found.  We are encouraged
that what we describe in this paper contains several ``next
steps'', with no dead end in sight.  There is at least one avenue
we can pursue further, but which lies beyond the scope of this
paper.

The direction we go will turn out to be close to the minimal
standard model. Our takeoff point is a fresh look at the mass
matrices of the quarks. The starting assumption is that the off
diagonal elements of the mass matrices are small relative to
diagonal elements (this in particular disallows the ``Fritzsch
texture" \cite{ref:10}). With this choice we search for simple
patterns. The candidate pattern which we pursue is that all
off-diagonal elements reside in the down-quark mass matrix.  This
leads to a relatively comfortable, but not quite compelling
``phenomenology", one element of which is that the unitarity
triangle may contain a right angle. Our next step is a very
natural one, namely to assume that this pattern is created by
mixing of ordinary down-quarks with three generations of heavy,
electroweak-singlet down-quarks \cite{silvermannir}. This is
naturally implemented in an anomaly-free way by extending each
generation of ordinary fermions to a 27-plet of $E(6)$
\cite{ref:11}.

In order to implement the generation of mass of the light quarks,
the right-handed heavy down-quarks are coupled to the ordinary
left-handed light-quark electroweak doublets and to the Higgs
bosons in a flavor diagonal way.  We assume that at least one of
these new Higgs couplings is as large as the top-quark Yukawa
coupling. If this is true for all three generations, then there is
enough modification in the evolution of these coupling constants
and of the Higgs self-coupling that there may be a strong-coupling
regime, along the lines of top-condensate models \cite{ref:12}, at
an oasis mass of order 1000 TeV. However, if only the new
third-generation Higgs coupling has this property, this need not
be the case. There is also an interesting modification of the
Higgs-sector effective potential which allows a novel, radiatively
generated mechanism for spontaneous symmetry breakdown. Finally,
because all flavor-changing effects originate in semi-soft mass
terms, it follows that all flavor-changing processes are finite
and calculable, with flavor change an ``infrared" phenomenon that
disappears at energy scales large compared to the mass scale of
the new heavy quarks and leptons.

All of these features we find not unwelcome.  In order for them to
occur, the masses of the new fermions should not exceed about 10
TeV. The lower limit comes from experimental constraints on
electroweak parameters and rare flavor-changing processes, while
the upper limit comes from a naturalness criterion on the
structure of the Higgs-sector effective potential.

In Section 2 we describe our approach to the mass-matrix problem.
In Section 3 the new heavy $E(6)$ fermions are introduced and the
Lagrangian for the standard-model extension is constructed. In
Section 4 we determine the phenomenological constraints. Section 5
examines the evolution of coupling-constants and the properties of
the possible oasis created by their evolution into a strong
coupling regime. There are opportunities for pursuing this line
further, and Sections 6  and 7 describe them and summarize the
situation.

\section{Is all Mixing in the Down Sector?}

One apparent feature of the CKM matrix is that all off-diagonal
elements are small compared to the diagonal ones. So it would seem
a very natural hypothesis that the same should be true for the
up-quark and down-quark mass matrices, namely that the
off-diagonal elements are small relative to the diagonal elements,
small enough to justify the use of perturbation theory.

This is not the most popular choice, however. More common is the
quite well-motivated one of ``Fritzsch texture", which  does not
allow the use of low order perturbation theory. Nevertheless, we
here pursue the perturbative option and see whether there is a
reasonably simple and credible scenario that gives a direct
relation between the CKM \cite{ref:13,ref:14} elements and
mass-matrix elements. We will find as candidate the aforementioned
option of putting all mixing in the down sector. We further find
that if the mass matrix has the form
\begin{equation}
\M_{ij} = m^{(S)}_i \delta_{ij} + i\epsilon_{ijk}m_k^{(A)}
\label{eq:2ai}
\end{equation}
with $m^{(S)}$ and $m^{(A)}$ real, then the unitarity triangle
turns out to good accuracy to be a right triangle. This structure
of mass matrix and CKM matrix is due to Stech \cite{ref:15}.

The mass terms for fermions are generated in the standard model by
the Yukawa coupling  term in the Lagrangian in the weak
interaction (primed) basis
\begin{equation}
\L = \L_{\rm up}+\L_{\rm down}
\label{eq:2a}
\end{equation}
where, \eg\
\begin{eqnarray}
\L_{\rm down} &=& \bar q'_{Li}\Phi h_{ij}q'_{Rj} +\hc \,,
\label{eq:motation} \\
         &\to& \bar q'_{Li}M'_{ij}q'_{Rj} +\hc \,, \nonumber
\end{eqnarray}
where $i$ and $j$ are flavor indices and $h_{ij}$ are coupling
constant matrices. Once the Higgs field $\Phi$ obtains a vacuum
expectation value and we shift the field $\Phi \to
(v+\phi)+i\mbox{\boldmath$\tau$\unboldmath} \cdot {\bf w}$, we
observe that $M'=hv$. (${\bf w}$ are the Goldstone fields eaten by
the $W$ and $Z$.) The mass matrix $M'$ can be diagonalized by a
bi-unitary transformation
\begin{equation}
M'=V_L~ M~V_R^{\dagger} \, .
\end{equation}
In the standard model after diagonalization, the only observable
relics of the individual mixing  matrices appears in the
left-handed charged current couplings of the  $W$ boson, since
right-handed charged currents are phenomenologically absent. Thus
\begin{equation}
V^{CKM}=V_L^{u\dagger}V_L^{d} \, .
\label{eq:2c}
\end{equation}
For $i\neq j$, this means that
\begin{eqnarray}
V^{CKM}_{ij} &=& \sum_k (V_L^{u})^{*}_{ki}(V_L^d)_{kj} \,,
\label{eq:2d}\\
             &\simeq& (V_L^{u})^{*}_{ji}+(V_L^d)_{ij}+(V_L^{u})^{*}_{\ell i}
(V_L^d)_{\ell j} \,, \nonumber
\end{eqnarray}
where now $i\neq j \neq \ell \neq i$ and the diagonal CKM elements
have been set to unity.

Let us assume that the original matrix $M'$ is hermitian; within
the standard model this can always be made true without any effect
on phenomenology by a judicious choice of $V_R$.  In this case,
the mass matrix is diagonalized by a simple unitary transformation
$V$. In the spirit of perturbation theory,  decompose the original
mass matrix $M'$ as
\begin{equation}
M'=M_0+m \,,
\label{eq:2e}
\end{equation}
where $M_0$ is diagonal and $m$ is purely off-diagonal. Then it is
straightforward to see that
\begin{equation}
M'~V=(M_0+m)V=VM \,,
\label{eq:2f}
\end{equation}
where $M$ is the diagonalized matrix. Since all these matrices are
assumed to be 3$\times$3, we can iterate this equation to obtain
an exact relationship between the elements of $V$ and those of
$M$, $M_0$ and $m$:
\begin{equation}
V_{ij} =
\left[\frac{m_{ij}}{(M_j-M_{0i})}+\frac{m_{ik}m_{kj}}{(M_j-M_{0i})
(M_j-M_{0k})}\right] V_{jj}
+\frac{m_{ik}m_{kj}V_{ij}}{(M_j-M_{0i})(M_j-M_{0k})} \,,
\label{eq:2g}
\end{equation}
with $i\neq j\neq k \neq i$ and no sums performed. This relation
can be directly solved for the ratio of $V_{ij}/V_{ii}$:
\begin{equation}
\left[\frac{V_{ij}}{V_{ii}}\right]^2
    = \frac{\alpha_{ij}}{\alpha^*_{ij}}
\left[\frac{M_j-M_{0j}-\frac{m_{jk}m_{kj}}{M_j-M_{0k}}}
       {M_j-M_{0i}-\frac{m_{ik}m_{ki}}{M_j-M_{0k}}}\right] \,,
\label{eq:2h}
\end{equation}
where again no summation is performed and
\begin{equation}
\alpha_{ij}=m_{ij}+{m_{ik}m_{kj}\over {M_j-M_{0k}}} \,,
\label{eq:2i}
\end{equation}
Note that the prefactor in the expression above,
$\alpha/\alpha^*$, is a pure
phase.

With these general preliminaries we may now address the problem at
hand. It is easiest to simply evaluate the relevant matrix
elements. We have to good approximation
\begin{eqnarray}
V_{12} &\cong&
\frac{m_{12}}{M_2} -
\frac{m_{13}m_{32}}{M_2M_3}
\cong
\frac{m_{12}}{M_2} \nonumber \\[2ex]
V_{23} &\cong&
\frac{m_{23}}{M_3} +
\frac{m_{21}m_{13}}{M^2_3}
\cong
\frac{m_{23}}{M_3}
\label{eq:2j}\\[2ex]
V_{13} &\cong&
\frac{m_{13}}{M_3} +
\frac{m_{12}m_{23}}{M^2_3}
\cong
\frac{m_{13}}{M_3} \ .\nonumber
\end{eqnarray}
The expressions for the conjugate elements $V_{31}$ and $V_{32}$
have larger second-order contributions, by a factor of order
$M_3/M_2$, and they are consistent with the unitarity constraints,
which are approximately
\begin{equation}
0 = [V^\dagger V]_{ij} \cong
V_{ij} + V^*_{ji} + V_{ki}^* V_{kj} \qquad
(i \ne k \ne j \ne i) \ .
\label{eq:2k}
\end{equation}
We have also seen in Eq. (\ref{eq:2d}) a similar phenomenon
occurring when combining together the expressions for $V^u$ and
$V^d$ into $V_{CKM}$: Most of the matrix elements are arguably
dominated by the first two additive terms in Eq. (\ref{eq:2d});
only large, unnatural cancellations between up and down rotations
could spoil the argument. In particular this situation rather
easily holds for $all$ the off-diagonal elements with the
exception of $V_{13}^{CKM}$ and $V_{31}^{CKM}$.

Thus, more concretely, we have
\begin{eqnarray}
V^{CKM}_{12} &\equiv& V_{us} \cong
\frac{m_{ds}}{M_s} +
\frac{m^*_{cu}}{M_c} \nonumber \\[2ex]
V^{CKM}_{23} &\equiv& V_{cb} \cong
\frac{m_{sb}}{M_b} +
\frac{m^*_{tc}}{M_t}
\label{eq:2l}\\[2ex]
V^{CKM}_{13} &\equiv& V_{ub} \cong
\frac{m_{db}}{M_b} +
\frac{m^*_{tu}}{M_t} -
\frac{m_{sb}}{M_b} \cdot
\frac{m^*_{cu}}{M_c} \ . \nonumber
\end{eqnarray}
Upon putting numbers into the above equations, there emerges the
vague outlines of a possible pattern, namely that all
non-vanishing off-diagonal elements of the mass matrix are in the
down sector, and that they are of the same order of magnitude,
essentially the same as the QCD confinement parameter
$\Lambda_{QCD}$. The magnitudes of the elements are as follows:
\begin{equation}
\left|M^\prime_{ij}\right| \cong \left(\begin{array}{ccr}
? & 35\ MeV & 15\ MeV \\
? & 150\ MeV & 175\ MeV \\
? & ? & 4.3\ GeV\end{array}\right) \ .
\label{eq:2m}
\end{equation}
The opposite extreme, all mixing in the up sector, evidently does
not possess this property.

It is also interesting, but not essential, to go one step further
and assume that the off-diagonal elements of the down-quark mass
matrix are imaginary antisymmetric, and that the diagonal elements
are real, as in the form of Eq. (\ref{eq:2ai}), which intuitively
does not appear unrealistic. Then we find the result, due to
Stech, that the unitarity triangle is a right triangle, with the
right angle being at the lower left ($\gamma = \pi/2$). This
result is a consequence of the fact that $V_{ub}$ is well
approximated by first order perturbation theory, hence pure
imaginary, and that the factors $V_{cd}^*$ and $V_{cb}$, whose
product form the base of the unitarity triangle, are each
well-approximated by the first order perturbation terms. Hence the
base of the unitarity triangle is real. Therefore the right angle
occurs, and in the right place. We note that this is not the same
argument as used recently by Fritzsch \cite{ref:10}, who finds the
upper vertex of the triangle to have the right angle ($\alpha =
\pi/2$).

The other two angles $\alpha$ and $\beta$ in the unitarity
triangle are determined as follows:
\begin{equation}
\alpha = \tan^{-1} \left(\frac{m_{sb}m_{ds}}{m_s m_{db}}\right) =
\left(\frac{\pi}{2}-\beta\right) \ . \label{eq:za}
\end{equation}
The values of $M^\prime_{ij}$ are not very precise.  They give a
range for $\alpha$ between 62$^\circ$ and 70$^\circ$, and for
$\beta$ between 28$^\circ$ and 20$^\circ$.  This implies that
$\sin 2\alpha = \sin 2\beta$ lies in the range 0.64 to 0.80.  The
value of $\beta$ agrees well with the recent measurements by BELLE
and BaBar \cite{ref:A}.

Of course this line is quite speculative, and not required. But
the data thus far is consistent with this option \cite{ref:16},
and if that remains the case when precision is increased, it might
be taken as {\it a posteriori} evidence for the credibility of the
line of attack we take.

\section{Heavy Fermions in the Tree Approximation}

According to the discussion in the previous section, we have been
motivated by simplicity and by the data to consider a scenario
where all flavor and CP violation originates in the down-quark
sector, perhaps via a mass term with the structure
\begin{equation}
\M_{ij} = \delta_{ij}m^{(S)}_i+i\epsilon_{ijk} m^{(A)}_k \ .
\label{eq:3a}
\end{equation}
If this is accepted, then the unitarity triangle is to good
approximation a right triangle, with $\gamma = 90^\circ$, as shown
in Fig. 1.

\vspace{.5cm}
\begin{figure}[htb]
\centering{\includegraphics{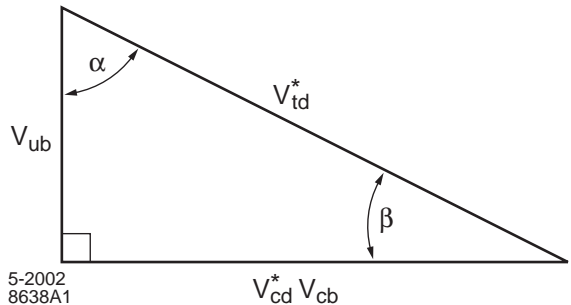}} \caption[*]{The right
unitarity triangle. \label{fig1}}
\end{figure}

In any case, we inquire here what dynamical mechanism might be
responsible for such a situation. There is a very natural answer,
namely that the ordinary down-quarks $d_i = (d,s,b)$ mix with
heavy down-quarks $D_i = (D,S,B)$ which are electroweak singlets
and have large intrinsic mass terms $M_i$. There is no mass mixing
in the up sector simply because the corresponding heavy up degrees
of freedom do not exist.

Such heavy down states are not altogether unwelcome. In GUT
extensions which go by way of $E(6)$, the fundamental
$\underline{27}$ representation contains just such a multiplet.
The $SU(5)$ decomposition of this $E(6)$ $\underline{27}$ is
\begin{equation}
27 = (\bar{10}+5)+(5+\bar 5)+(1+1) \ .
\label{eq:3b}
\end{equation}
We notice that if this is the scenario, then there will also be
heavy leptons $L_i = (E,M,T)$ which are vector electroweak
doublets, \ie\ both the left-handed and right-handed components of
the $L_i$ couple to the $W$'s. We shall here accept that this is
the case, because even without $E(6)$ one would have a problem
with anomaly cancellation if only the heavy down-quarks $D_i$ were
introduced. Adding in the $L_i$ clearly solves the anomaly
problem.

The mass mixing term for the down-quarks, as shown in Fig. 2
(consistent with $SU(2)\times U(1)$ electroweak symmetry), now can
be written down:
\begin{equation}
\L^\prime_\mu = \sum_{ij} \bar D_{iL}\, \mu_{ij}\, d_{jR} + \hc
\label{eq:3c}
\end{equation}
with, perhaps, $\mu_{ij}$ having the form of Eq. (\ref{eq:3a}). In
addition we must introduce intrinsic mass terms (taken to be real)
for the heavy down-quarks (also consistent with electroweak gauge
symmetry)
\begin{equation}
\L^\prime_M = \sum_i M_i \bar D_{iL}D_{iR} + \hc \label{eq:3e}
\end{equation}
This much does not by itself give mass to the light down-quarks,
because their left-handed components are as yet not coupled. This
is remedied by assuming that the Higgs bosons couple the heavy
right-handed down-quarks to the left-handed ordinary quarks.
\begin{equation}
\L^\prime_H = \sum_i \bar q_{iL}\Phi H_iD_{iR} + \hc
\label{eq:3f}
\end{equation}
where
\begin{equation}
q_i \equiv {u_i\choose d_i} \qquad \Phi = (v+\phi)+
i\mbox{\boldmath$\tau$\unboldmath} \cdot {\bf w}  \label{eq:3g}
\end{equation}
and
\begin{equation}
H_i  \equiv \left( \begin{array}{cc} 0 & 0 \\ 0 & H_i
\end{array}\right)\ . \label{eq:3h}
\end{equation}
Then the mass matrix of the light quarks is obtained by
diagonalization, essentially second-order perturbation theory:
\begin{equation}
m_{ij}\cong (H_iv)\ \frac{1}{M_i}\, \mu_{ij}\ .
\label{eq:3i}
\end{equation}
with our convention that
\begin{equation}
\L^{(\rm effective)}_m = \bar d_{iL}\, m_{ij}\, d_{jR} + \hc
\label{eq:3j}
\end{equation}

\vspace{.5cm}
\begin{figure}[htb]
\centering{\includegraphics{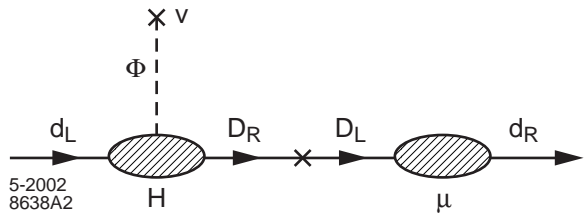}} \caption[*]{Mixing
mechanism for the down-quark masses. \label{fig2}}
\end{figure}

\noindent Note that we take care to omit the usual Higgs coupling
between the right-handed ordinary down-quarks $d_i$ and the
ordinary left-handed quarks. This assumption is robust in the
sense that, if the terms in the mass matrix which mix $d$ and $D$
are neglected, the full Lagrangian as written possesses enough
symmetry that such Yukawa terms will not be generated by radiative
corrections. This conclusion follows immediately from the fact
that in that limit the right-handed $d$'s decouple completely from
the Higgs sector. In other words, the number of $d_R$'s of each
generation is, in the absence of the mass-mixing term, Eq.
(\ref{eq:3c}), conserved. Inclusion of the kinetic energy terms
and gauge couplings still leaves the Lagrangian invariant under
independent phase transformations of the $d_R$'s. Therefore for
small $\mu$ and $ m$, as we shall see, the $\mu$ dependent terms
are finite and calculable.

We will need to include in the Lagrangian the usual Higgs
couplings of the right-handed up-quarks $u_i = (u,c,t)$ to their
left-handed counterparts.  However recall again that these
couplings are assumed to be flavor diagonal. Just to set the
notation, the Lagrangian for these terms is
\begin{equation}
\L^\prime_h = \sum_i \bar q_{iL}\, \Phi\, h_i\, u_{iR} \qquad
h_i \equiv \left(\begin{array}{cc} h_i & 0\\ 0 & 0\end{array}
\right) \ .
\label{eq:3k}
\end{equation}
We should also include the gauge boson terms as well.  However,
for almost all purposes in this paper, we can neglect their
presence, remembering that in the limit of vanishing gauge
couplings the longitudinal $W$ and $Z$ become the massless
Goldstone modes of the Higgs sector.

It is important to estimate as well as possible the magnitudes of
the new parameters. The new Yukawa couplings $H_i$ may be rather
large, of order of the top quark coupling $h_t$ or larger, and as
we shall see this case appears to be  an interesting one. The
magnitudes of the masses of the new down-quarks and leptons will
be constrained from below by the precision electroweak data and
perhaps from above by the properties of the Higgs-potential
radiative corrections. The nominal values are in the 1--10 TeV
regime. The constraints on the parameters will be considered in
more detail in the next section.

However before turning to that, we are ready to diagonalize the
mass matrix of the quarks.  We write the 6$\times$6 undiagonalized
mass matrix $\M^\prime$ in block form as
\begin{equation} \M^\prime =
\left(\begin{array}{cc} 0 & Hv \\ \mu & M\end{array} \right)\ ,
\label{eq:3l}
\end{equation}
where the indices 1,2,3 label the ordinary light quarks $d_i$ and
4,5,6 the new heavy quarks $D_i$. This matrix is not hermitian, so
we diagonalize its square $\M^\prime \M^{\prime\dagger}$, because
it is the left-handed mixing which we need to see explicitly in
the CKM phenomenology.  Note that $H$ and $M$ are 3$\times$3
diagonal matrices, so that we may write
\begin{equation}
\M^\prime \M^{\prime\dagger} = \left(\begin{array}{cc} H^2v^2 & HvM\\
HvM & \mu\mu^\dagger + M^2\end{array}\right) = \V\M^2\V^\dagger \
, \label{eq:3m}
\end{equation}
where $\M$ by definition is diagonal, and we are to find $\V$.  A
first prediagonalization is immediate, if one ignores the (small)
terms $\mu$ associated with the flavor and CP violation.  In this
approximation, the diagonalizing matrix $\V$ is 3$\times$3 block
diagonal
\begin{equation} \V = \left( \begin{array}{cc}
c & s \\ -s & c \end{array}\right)
\label{eq:3n}
\end{equation}
where
\begin{equation}
s \cong \frac{Hv}{M} \equiv \left( \begin{array}{ccc}
\sin\theta_1 & 0 & 0 \\ 0 & \sin\theta_2 & 0 \\ 0 & 0 &
\sin\theta_3
\end{array}\right)
\label{eq:3o}
\end{equation}
and
\begin{equation} c^2
+ s^2 = 1 \ .
\label{eq:3p}
\end{equation}
We will find in the next section that these mixing angles cannot
be large.  The reason is simple; the electroweak-doublet nature of
the left-handed quarks is diluted by the mixing, and the $W$ and
$Z$ do not couple to the singlet portion with the same strength as
to the doublet portion. Therefore there are renormalizations of
the strengths of the diagonal couplings of $W$ and $Z$  to quark
pairs, and the precision electroweak data restricts these
renormalizations to be no larger than the 1\%\ level.  Since the
depletions are proportional to the square of the mixing angles,
this limits the $s_i$ to no more than the 10\%\ level.

Now let us examine in more detail the residual nondiagonal terms.
Thus far we have
\begin{eqnarray}
\V^\dagger(\M^\prime \M^{\prime\dagger})\V
&=& \left(\begin{array}{cc}0 & 0 \\ 0 &
M^2+H^2v^2\end{array}\right) +
\V^\dagger\left(\begin{array}{cc} 0 & 0 \\ 0 & \mu\mu^\dagger
\end{array}\right)
\, \V \nonumber \\ &=& \left(\begin{array}{cc}  s\mu\mu^\dagger s
& -s\mu\mu^\dagger c \\\ \  -c\mu\mu^\dagger s\ \ \ &
c\mu\mu^\dagger c +
M^2+H^2v^2 \end{array}\right) \ .
\label{eq:3q}
\end{eqnarray}
We see that, as expected, the mass matrix of the light quarks to
this order is
\begin{equation}
mm^\dagger \cong (s\mu)(s\mu)^\dagger
\label{eq:3r}
\end{equation}
with
\begin{equation} m = s\mu = \frac{Hv}{M}\ \mu
\label{eq:3s}
\end{equation}
consistent with Eq.  (\ref{eq:3i}). Without going further we see
from Eq.  (\ref{eq:3q})\ that if the $s_i$ are flavor universal,
then diagonalization with the CKM matrix suffices to diagonalize
to all orders the full mass matrix.  However, this is not true in
general.  More generally, the next-order correction to the light
quark mass matrix has the form

\begin{eqnarray}
\Delta(mm^\dagger) &=& (s\mu\mu^\dagger c) \
\frac {1}{M^2+H^2v^2}\ (c\mu\mu^\dagger s) = (s\mu\mu^\dagger s)\
\frac{c^2}{s^2(M^2+H^2v^2)}\ (s\mu\mu^\dagger s) \nonumber\\
&=& (s\mu\mu^\dagger s)\
\left[\frac{1}{H^2v^2}-\frac{1}{(M^2+H^2v^2)}\right]\
(s\mu\mu^\dagger s) \ .
\label{eq:3t}
\end{eqnarray}

In any case, the flavor- and CP-violation effects are concentrated
in a semisoft mass term. This in turn leads to only small, finite,
calculable corrections to the Yukawa couplings and heavy-quark
mass terms.  As will be demonstrated in more detail in the next
section, flavor violation in this model becomes an ``infrared''
phenomenon only. Flavor nonuniversality, \ie\ diagonal
flavor-conserving couplings depending upon the generation, will
however persist more than does the flavor violation.  But these
will be small if the $H_i$ and $M_i$ are flavor universal.  (Of
course the up-quark couplings $h_i$ cannot be flavor-universal,
and that nonuniversality alone suffices to create corrections in
all the other couplings via radiative effects.)

In the lepton sector, things are slightly different.  The
left-handed charged leptons, $\ell_L \equiv (e, \mu, \tau)$ mix
with the heavy leptons, $ L^- \equiv (E,^-M,^-T^-)$ via a
semi-soft mass term $\widetilde{\mu}$ like the d-quarks while the
right-handed ones $\ell_R$ couple to $L^-_L$ via the Higgs Yukawa
couplings yielding a 6$\times$6 mass matrix of the form
\begin{eqnarray}
\M_\ell =
\left (
\begin{array}{cc}
0   &  \widetilde{H} v \\
\widetilde{\mu} & \widetilde{\M}
\end{array}  \right ) \ .
\end{eqnarray}

Just  as for the quarks, there is again no Yukawa coupling term
which would couple the ordinary light leptons $\ell_L$ to the
$\ell_R$.  This time a nonvanishing  additive quantum number
assigned to the ordinary left-handed lepton electroweak doublet
$(\ell, \nu_\ell)_L$ and to nothing else suffices, in the absence
of the flavor-changing mixings proportional to $\mu$, to forbid
the presence of this Yukawa coupling.  Alternatively, the
left-handed lepton doublet might be assigned a multiplicative
``parity" quantum number -1; with all other multiplets having
positive parity.  Note that this assignment survives the extension
to the GUT $SU(5)$ level.

In the neutrino sector, the mixing of $\nu^i_L$ with the $E^0_L$
leaves the neutrinos massless.  But the mixing of $\nu_L$ with the
$SU(5)$-singlet fermions present in the $E(6)$ extension will give
rise to neutrino masses and mixings.  In principle we need not
introduce these extra degrees of freedom $(N,N')$ at all. But
there is ample motivation for doing so, and we return in the next
section to a discussion of the implications.

The upshot of all this is that there is a 6$\times$6 unitary
matrix diagonalizing the charged lepton mass matrix with a
structure very similar to the one for down-quarks:
\begin{equation}
\widetilde{U}_6 = \left (
\begin{array}{cc}
\widetilde{U}_3 & 0 \\
0 & 1 \\
\end{array} \right )
\left (
\begin{array}{cc}
\widetilde{c} & \widetilde{s}  \\
-\widetilde{s} & \widetilde{c}
\end{array}   \right )
\end{equation}
with $\widetilde{s}_i \approx \widetilde{H}_i v / \widetilde{M}_i$
etc. The feature of nonuniversality of diagonal couplings will
persist, and the phenomenology differs slightly due to the fact
that the heavy leptons are vector-like electroweak doublets, not
singlets.

Before going on, we summarize how the tree level couplings to $W $
and $Z$ are modified by the mixing effects. For the $W$, there
will be the usual CKM structure, but modified by the effects of
the mixing to the heavy down-quarks expressed by $\V$. Thus
\begin{equation}
G_F \bar u_{iL}\gamma^\mu(V_{CKM})_{ij}d_{jL} \Rightarrow G_F\,
\left \{ \bar u_{iL}\gamma^\mu (V_{CKM})_{ij}c_jd_{jL} \  +
\bar{u}_{iL}  \gamma^\mu (V_{CKM})_{ij} s_j   D_{jL}  \right \}.
\label{eq:3u}
\end{equation}
In the same way the leptonic coupling strengths to the $W$ are
modified:
\begin{equation}
G_F \bar \nu_{iL}\gamma^\mu\ell_{iL} \Rightarrow G_F \left \{ \bar
\nu_{iL}\gamma^\mu (V_{MNS})_{ij} \widetilde c_j\ell_{jL} +
\bar{\nu}_{iL} \gamma^\mu (V_{MNS}){_{ij}} \widetilde{s}_j
L^-_{jL} - \bar{L}^0_{iL} \gamma^\mu \widetilde{s}_{i} \ell_{iL}
\right \} \label{eq:3v}
\end{equation}
where the leptonic mixings $\widetilde s_i, \widetilde c_i$, are
introduced in analogy to the quark mixings $s_i,c_i$ (Eq.
(\ref{eq:3n})), and in general are different. The leptonic mixing
matrix $V_{MNS}$ (named after Maki, Nakagawa and Sakata
\cite{ref:17}) is the analog of the CKM matrix in the quark
sector. It is given by
\begin{equation}
V_{MNS} = U^\dagger_\nu \ \widetilde  U_3 \ , \label{eq:zb}
\end{equation}
where $U_\nu$ is the matrix which diagonalizes the neutrino mass
matrix.
 In addition, there are new
right-handed leptonic charged currents given by:
\begin{equation}
 \L^\prime_L =G_F \overline{L}^0_{iR}  \  \gamma^\mu   (-\widetilde{s}_i )
 \ell_{iR} \ .
\end{equation}
The couplings of light down-quarks to the $Z$ are affected.
Starting from the usual structure
\begin{equation}
 \L_{NC} \propto T_3 - \sin^2\theta_WQ \ ,
\label{eq:3w}
\end{equation}
only the $T_3$ portion will be affected by all this mixing; charge
conservation protects the rest. The effect of the heavy-light
mixing is, for the quarks,
\begin{equation}
(\bar u_{iL}\gamma^\mu u_{iL}-\bar d_{iL}\gamma^\mu d_{iL})
Z^\mu \Rightarrow (\bar u_{iL}\gamma^\mu u_{iL} -
c^2_i\bar d_{iL}\gamma^\mu d_{iL})Z_\mu - c_is_i \bar{d}_{i L}
\gamma^\mu
D_{iL} Z_\mu + h.c.
\label{eq:3x}
\end{equation}
For the leptons,
\begin{equation}
(\bar\nu_i\gamma^\mu\nu_i - \bar \ell_{iL}\gamma^\mu \ell_{iL})
Z^\mu \Rightarrow (\bar\nu_i\gamma^\mu\nu_i
-\bar\ell_{iL}\gamma^\mu\ell_{iL} -
\widetilde{s}_i^2\bar\ell_{iR}\gamma^\mu \ell_{iR} +
\widetilde{s}_i \widetilde{c}_i \bar{\ell}_{iR} \gamma^\mu L_{iR})
Z^\mu \ . \label{eq:3y}
\end{equation}

The phenomenological implications of these and other modifications
will be taken up in the next section.

\subsection{Neutrino Masses}

Neutrino mixing with $E^0$ gives rise to a mass matrix (for each
flavor)
of the form:
\begin{eqnarray}
\left (
\begin{array}{ccc}
0      &    m  \\

0      &    M
\end{array} \right ) \ .
\end{eqnarray}
This has one zero eigenvalue and so the neutrinos remain massless.
When one of the two gauge singlets, $N$, is included, the mass
matrix becomes:
\begin{eqnarray}
\left (
\begin{array}{ccc}
0      & m     & m'  \\
0      & M     & 0   \\
m'     & 0     & M'
\end{array} \right ) \ .
\end{eqnarray}
The neutrino mass eigenvalue is now given by the usual see-saw
formula:
\begin{eqnarray}
m_\nu \sim m'^2/M' \ .
\end{eqnarray}
If the Dirac masses $m'$ are taken to be of the order of the
top-quark mass, then the mass scale $M'$ lies in the range of
$10^{14}$ GeV in order to accommodate the typical mass scales
required in the atmospheric and solar neutrino oscillations. This
is lower than the unification scale, and might represent an oasis.
The other singlet $N'$ might be at the unification scale. To make
some ansatz about the flavor structure of the lepton and neutrino
mass matrices, which would lead to a desirable form for the MNS
matrix, is beyond the scope of this paper and we leave it to
another occasion.
\section{Phenomenological Implications}

\subsection{Decay Widths}

The masses of the new heavy quarks and leptons are free parameters
in this model. As we shall soon see, unless the Higgs couplings
$H_i$ are small, there are difficulties with precision electroweak
data if these masses are low. The experimental constraints will,
as discussed below, generically lead to bounds on the mixing
angles $s_i$ and $\widetilde s_i$, as defined in Eq.
(\ref{eq:3o}):
\begin{equation}
| s_i| = \left|\frac{H_iv}{M_i}\right| \lsim 0.1 \ . \label{eq:zc}
\end{equation}
If the masses are large in comparison
with the gauge boson and even Higgs masses, the decay
phenomenology of the new heavy fermions is quite straightforward:
they will decay predominantly into the ordinary quark or lepton of
the same generation together with either a Higgs or a
(longitudinal) gauge boson. For example, for the third generation
$B$ we have the decay modes
\begin{eqnarray}
B &\rightarrow& b + h \\[1ex]
B &\rightarrow& b + Z^0 \\[1ex]
B &\rightarrow& t + W^- \end{eqnarray}
with
\begin{equation}
\Gamma(B \rightarrow bh) = \Gamma(B\rightarrow bZ^0) =
\frac{1}{2}\, \Gamma(B \rightarrow tW^-) \ .
\label{eq:aa}
\end{equation}
The total decay width is, in the limit of negligible final-state
masses,
\begin{equation}
\Gamma(B\rightarrow {\rm all}) \cong \frac{H^2_B}{4\pi}\ M_B \ .
\label{eq:ab}
\end{equation}
The main usefulness of this formula is to provide a practical
upper bound on the magnitude of the Higgs couplings $H_i$; the
theory makes little sense if the width of the parent fermion is
larger than its mass. In particular this constraint implies
\begin{equation}
|H_i| \leq 4
\label{eq:ac}
\end{equation}
or
\begin{equation}
\frac{H^2_i}{4\pi} \lsim 1 \ .
\label{eq:ad}
\end{equation}
The decay phenomenology for heavy leptons is essentially
identical; the decay modes for the third generation $T$ are given
by:
\begin{eqnarray}
T^- \longrightarrow \tau^- + Z^0 \nonumber \\
T^- \longrightarrow \tau^- + h^0   \\
T^0 \longrightarrow \tau^- + W^+  \nonumber
\end{eqnarray}
and the widths satisfy
\begin{equation}
\Gamma (T^- \rightarrow \tau^- h^0)  = \Gamma (T^- \rightarrow
\tau^-Z^0 ) = \frac{1}{2} \Gamma (T^0 \rightarrow  \tau^- W^+) \ .
\end{equation}
The total widths are given by:
\begin{equation}
\Gamma_{tot} (T^- ) = \Gamma_{tot} (T^\circ) =
\frac{\widetilde{H_T}^2 \widetilde{M_T}}{8 \pi}
\end{equation}
and thus $\widetilde{H}_i$ satisfy bounds similar to $H_i$.

In the above scenario, the search strategies for these particles
are in principle straightforward.  The main issue is attaining
sufficient energy to produce them.

On the other hand, if some of the new Yukawa couplings $H_i$ are
small, then the associated masses can be small as well. However,
we shall consider it unreasonable that {\it all} the new quarks
have Yukawa couplings to the Higgs sector much smaller than the
coupling of the top quark to the Higgs.  In particular we
hereafter assume that the third generation down-quark $B$ has a
Yukawa coupling at least as large as that of the top quark:
\begin{equation}
H_B \geq h_t \cong 0.70 \ .
\label{eq:zd}
\end{equation}
We shall see later that this innocent hypothesis leads to
interesting consequences.

Within this scenario there are still major choices to make.  One
extreme is to assume flavor symmetry for the new heavy down-quark
multiplet:
\begin{eqnarray}
M_B \approx M_S \approx M_D \nonumber \\
H_B \approx H_S \approx H_D \nonumber .
\label{eq:ze}
\end{eqnarray}
The other extreme is to assume a mass and coupling
constant hierarchy as found everywhere else in the spectrum of
fermion degrees of freedom:
\begin{eqnarray}
M_B \gg M_S \gg M_D \nonumber \\
H_B \gg H_S \gg H_D \ . \label{eq:zf}
\end{eqnarray}
While many variants can be entertained, in what follows we
restrict our attention to these two extremes.  In particular, in
the hierarchial choice, we find no reason why the first and second
generation heavy quarks and leptons cannot be as light as given by
the present direct experimental limit on their production, about
130 GeV \cite{ref:x}. If such masses do approach the direct
experimental limits for production of heavy quarks and leptons,
the phenomenology may differ significantly. However, one must keep
in mind that a large violation of universality of the new
parameters $H_i$ and $M_i$ may be constrained by precision
electroweak data. These are discussed in the following
subsections.

\subsection{Tree-Level Mixings}

In the previous section we have exhibited the effective Lagrangian
obtained by diagonalization of the mass matrix of the light
quarks. In addition to leaving behind the CKM mixing of the quark
couplings to the $W$ bosons, we have seen there are nonuniversal
diagonal couplings to the $Z$ boson, which lead to deviations of
the values of electroweak parameters from the standard model
values. The percentage deviations will be proportional to the
squares of the mixing angles $s_i$ introduced in Eq. (30). For
example, the branching fraction $R_b$ for $Z$ decay into $\bar b
b$ is modified as follows
\begin{equation}
R_b \rightarrow R_b^{SM}
\left[1-\frac{2s^2_{3}\left(1-\frac{2}{3}\sin^2
\theta_W\right)}{\left(1-\frac{4}{3}\sin^2\theta_W + \frac{8}{9}
\sin^4\theta_W\right)}\right]
\label{eq:ae}
\end{equation}
with similar expressions, of course, for decays into $\bar s s$ or
$\bar d d$. Likewise the forward-backward asymmetry $A_{FB}$ in
$Z$ decays to $\bar b b$ is modified as follows:
\begin{equation}
(A_{FB})_b \rightarrow (A_{FB})_b^{SM}\,
\left[1-\frac{\frac{16}{9}\, s^2_{3}\sin^4\theta_W
\left(1-\frac{2}{3}\sin^2\theta_W\right)}
{\left(1-\frac{2}{3}\sin^2\theta_W)^4 - (\frac{2}{3}
\sin^2\theta_W\right)^4}\right]  \ .
\label{eq:af}
\end{equation}
Modifications to the diagonal couplings of the $Z$ to leptons is
also of significance. The most sensitive constraint comes from
measurements of the polarization asymmetry $A_{LR}$. The
correction modifies $A_{LR}$ as:
\begin{equation}
A_{LR} =
\frac{2(1-4\sin^2\theta_W)+8\sin^2\theta_W\widetilde{s}^2_1}
{1+(1-4\sin^2\theta_W)^2-8\sin^2\theta_W \widetilde{s}^2_1} \ .
\label{eq:ag}
\end{equation}
However, competitive tests also come from the measurement of the
axial couplings $g_A$ of the $Z$ to lepton pairs. The mixing
correction decreases $g_A$
\begin{equation}
g_{A_i} = (g_R-g_L)_i = (T_3\widetilde{s}^2_i-\sin^2\theta_WQ)
-(T_3-\sin^2\theta_WQ) = \frac{1}{2}\, (1-\widetilde{s}^2_i) \ .
\label{eq:ah}
\end{equation}

In all these cases the current data are consistent with the
standard model. From LEP and SLC data, deviations from the
standard model are bounded \cite{ref:19} and yield bounds in turn
on $s_3$ and $\widetilde{s}_1$ of about  0.1 and 0.05
respectively. Similar bounds can be deduced for $s_1$,
$\widetilde{s}_2$, and $\widetilde{s}_3$. We expect similar bounds
on the other mixing angles. The bound on $s_3$ comes from $R_b$.
Should the current discrepancy in $A_{FB}$ of $-0.0046 \pm 0.0017$
turn out to be real, we cannot account for it (an $s_3$ of 0.1
allows a maximum deviation of $-$0.001).

\subsection{Mixings at One-Loop}

\subsubsection{Flavor Mixing}

The modified charged currents in Eq. (38) and (39) have a very
special form.  In particular, even though the effective 3$\times$3
CKM (or MNS) matrix is not unitary, it has partial orthogonality
i.e.
\begin{eqnarray}
\sum_{i=u,c,t} \left ( V_{CKM} \right )_{bi}^* \left (V_{CKM}
\right)_{is}
 = c_3 c_2 \sum_{i} U_{3i}^* U_{i2} = 0
\end{eqnarray}
and similarly:
\begin{eqnarray}
\sum_{i=\nu_1,\nu_2,\nu_3} \left ( V_{MNS} \right )_{\mu i}^*
\left (V_{MNS} \right)_{ie}
 = 0  \ .
\end{eqnarray}
However, this is not true in the up-quark sector:
\begin{eqnarray}
\sum_{i=d,s,b} \left (V_{CKM} \right )_{ci}^* \left (V_{CKM}
 \right )_{iu} \neq 0 \ .
\end{eqnarray}

This means that there are no new one-loop flavor changing effects
in the down-quark sector or the lepton sector.  Hence, there are
no new contributions beyond the standard model in: (i) $\delta
m_K$, (ii) $\epsilon$, (iii) $\epsilon'/\epsilon$, (iv) $\delta
m_B$, (v) b $\rightarrow s \gamma$ or $b \rightarrow d \gamma$,
(vi) CP violating effects in the b-quark sector, (vii) $\mu
\rightarrow e\gamma$, (viii) $K_L \rightarrow \mu e$ etc.  This is
a strong prediction of this set of ideas.

There is a potential new contribution to $D^0-\overline{D^0}$
neutral charm meson mixing due to the heavy quarks in the box
diagram. Assuming for simplicity that the heavy $S$ quark
dominates, the new contribution to $\delta m_D$ is given by
\cite{ref:20}
\begin{eqnarray}
\delta m_D = \frac{G_F^2} {6 \pi^2} \  f^2_D \ B_D \ m_D  \ m^2_W
\ s_2^4 \left ( V^{*}_{cs} V_{su} \right ) ^2 \ f(x_S) \ .
\end{eqnarray}

For large $M_S, f(x_S) \approx M_S^2/4 m_W^2$ and $\delta m_D$
becomes:
\begin{eqnarray}
\delta m_D &=& \frac{G_F^2} {24 \pi^2} \  f^2_D \ B_D \ m_D  \
M^2_S \ s_2^4 \ \theta^2_c \\ \nonumber & =& (2\times 10^{-9}
s^4_2)\, GeV (M_S/1\ TeV)^2 \ .
\end{eqnarray}
If the heavy $S$ quark is replaced by $D$, a similar bound is
obtained.  If $S$ is replaced by $B$, the contribution is much
smaller.

The current bound on $\delta m_D$ is $5 \times 10^{-14}$ GeV and
demands $s_2 < 0.07$. This remains a potential significant
contribution to $\delta m_D$.

\subsubsection{S, T and U Parameters}

A convenient parametrization which describes new physics
contributions to electroweak radiative corrections is given by the
$S,T,U$ formalism of Peskin and Takeuchi \cite{peskin2}.  These
parameters are defined such that deviations from zero would signal
the existence of new physics (an alternate set of parameters,
$\epsilon_i$ also exist which do not require a reference point for
the standard model \cite{altarelli}).

The contributions from the extra fermions introduced here are
rather small as shown in Ref. \cite{hewett}. At a  mass scale of
$O(TeV)$ the contributions to $S,T$ and $U$ are no more than a few
times $10^{-3}$ and well within the current bounds, as long as the
heavy lepton doublets are nearly degenerate.

\subsubsection{Anomalous Magnetic Moment of the Muon}

\begin{figure}[htb]
\centering{\includegraphics[height=4in]{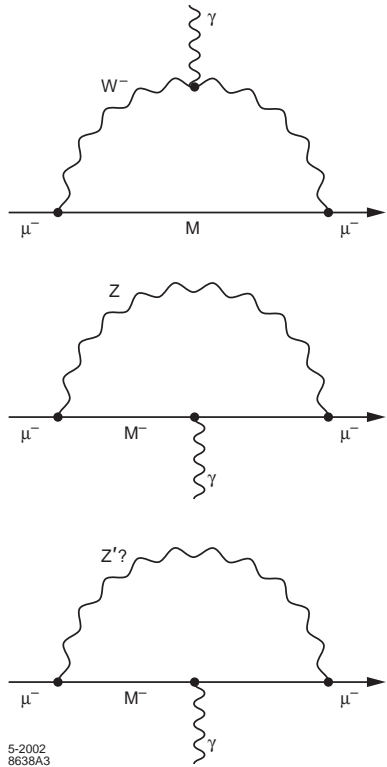}
\caption[*]{Feynman graphs for the ($g-2$) correction.}
\label{fig:3}}
\end{figure}

The extraordinary precision of the $g-2$ measurements suggests
that significant constraints on our mixing and mass parameters
might exist. With this in mind we consider the contribution of the
heavy leptons $M^-$ and $M^0$ to $a_\mu$.  There are two Feynman
diagrams as given in Fig.~\ref{fig:3}.   We let  $M_M^- \approx
M_M^0 \approx M_M$ each be about a TeV.  Then the total
contribution to $a_\mu = (g - 2)_\mu$ is given by \cite{leveille,
rizzo}.
\begin{equation}
a_\mu = \frac{G_F m^2_\mu}{4 \sqrt{2} \pi^2} ( \widetilde{s}_2)^2
\{ F + H\}
\end{equation}
where $F$ and $H$ are slowly varying functions of $(M_M/M_Z)^2$
and $(M_M/M_W)^2$ respectively \cite{leveille,rizzo}.  For the
value $M_M \sim 1\  TeV$,  the combination $\{F + H\} \approx +
1/3$ and $a_\mu$ is given by:
\begin{equation}
a_\mu \sim (\widetilde{s}_2)^2 \  74 \times 10^{-11}
\end{equation}
and for $\widetilde{s}_2 \sim$  0.1, this contributes to $a_\mu$
something less than $10^{-11}$.

\vspace{.5cm}

If the extra $Z'$ contained in $E(6)$ is relatively light, say of
order $1\ TeV$; then there is a potentially large contribution to
the $a_\mu$ from the Feynman diagram in Fig.~\ref{fig:3} where
both $M_M$ and $Z'$ appear in the loop.  The dominant contribution
to $a_\mu$ from this graph is given by \cite{leveille, rizzo}.
\begin{equation}
a_\mu = \frac{\alpha'}{2 \pi} \frac{m_\mu}{M_z'}^2
(\widetilde{s}_2)^2 \widetilde {G} (\frac{M_M}{m_\mu}) X
\end{equation}
where $X$ is given by:
\begin{equation}
X = \left[-
(x^\mu_L - x^M_L)(x^\mu_R -x_R^M) \right],
\end{equation}
$\widetilde{G}$ is a slowly varying function of $(M_M / M_Z')^2$,
and $(M_M/m_\mu)$ is the enhancement due to chirality violation.
For $(M_M/M_Z') \sim 1$, the function $\widetilde{G}$ is close to
1/2. The value of $X$ depends on the details of how the symmetry
is broken from $E(6)$ and can be as large as $+~18$ as shown in
Ref. \cite{hewett1}. Then, for $\alpha' \sim \alpha$,
\begin{equation}
a_\mu (Z') \sim  (\widetilde{s}_2)^2\  504 \times 10^{-11} \ .
\end{equation}
If $\widetilde{s}_2$ is about 0.1, then this contribution to
$a_\mu$ can be no more than $5\times 10^{-11}$ and is not
important.

We conclude that the sensitivity of the $g-2$ measurements appears
to be somewhat milder than the other precision electroweak
measurements we have considered.

\subsection{One-loop Modification of the Higgs Effective
 Potential}

The new one-loop contributions to the Higgs potential are of
considerable interest. Recall that the renormalized standard-model
effective potential can be written, up to a quadratic polynomial
in $\phi^2$, as
\begin{eqnarray}
V &\simeq& \frac{\lambda}{4} (\phi^2-v^2)^2 \nonumber \\[2ex]
&& +\  \frac{\lambda^2(3\phi^2-v^2)^2}{64\pi^2}\, \ell n\,
\frac{(3\phi^2-v^2)}{2v^2} \nonumber \\[2ex]
&& + \ \frac{3\lambda^2}{64\pi^2}\ (\phi^2-v^2)^2\, \ell n\,
\frac{(\phi^2-v^2)}{v^2} \\[2ex]
&& - \ \frac{12m^4_t}{64\pi^2}\ \frac{\phi^4}{v^4}\, \ell n\,
\frac{\phi^2}{v^2} \nonumber \\[2ex]
&& + \ \frac{3}{64\pi^2}\, (2m^4_W + m^4_Z)\,
\frac{\phi^4}{v^4}\, \ell n\,
\frac{\phi^2}{v^2} \ . \nonumber
\label{eq:52bb}
\end{eqnarray}
where the radiative terms are renormalized such that only the
logarithmic terms are kept and that at $\phi^2=v^2$ they vanish.
Note that in this scheme there will be residual finite radiative
corrections to the vacuum condensate strength $v^2$.

\vspace{.5cm}
\begin{figure}[htb]
\centering{\includegraphics[height=2.3in]{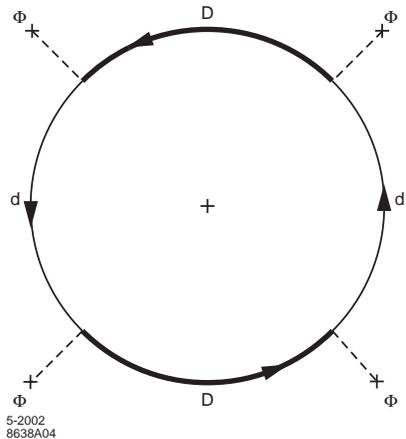}}
\caption[*]{Heavy down-quark contribution to the effective
potential. \label{fig:4}}
\end{figure}

The new contribution after renormalization has a very interesting
structure due to the fact that the new fermion loop has equal
numbers of heavy-quark segments and light-quark segments (Fig.
\ref{fig:4}):
\begin{equation}
\delta V = - \frac{12}{64\pi^2} \sum^3_{i=1}\,
\left(M^2_i+ H^2_i\phi^2\right)^2\, \ell n\,
\frac{M^2_i+H^2_i\phi^2}{M^2_i+H^2_iv^2} \ .
\label{eq:53aa}
\end{equation}
The important feature is that at scales below the heavy-fermion
mass scale $M$, there is a new (finite!)  quadratic mass term for
the Higgs lagrangian, with negative coefficient.
\begin{eqnarray}
\delta V &\cong& - \frac{12}{64\pi^2} \sum^3_{i=1}\, H^2_i (M^2_i
+ H^2_i v^2)^2 (\phi^2-v^2)/M_i^2 \nonumber \\
&\cong& -\frac{12}{64\pi^2} \sum^3_{i=1} H^2_i M^2_i (\phi^2-v^2)
. \label{eq:53bb}
\end{eqnarray}
This means that the spontaneous symmetry breaking can be induced
radiatively via these new heavy fermions.

One may question whether this conclusion is significant, since it
is an argument on the structure of the renormalization constant
$v^2$.  One might argue that a quadratic polynomial in $\phi^2$,
with coefficients dependent on $H_i$ and $M$, should be appended
to Eq. (\ref{eq:53aa}) such that, in the limit $M_i \rightarrow
\infty$, $\delta V$ vanishes.  This would be in accordance with
Appelquist-Carrazone \cite{ref:21} behavior---although the
arguments of Appelquist and Carrazone do not strictly apply to
this case. Evidently such a counterterm will eliminate the
quadratic term of interest in Eq. (\ref{eq:53bb}). But it seems
equally reasonable to take Eq. (\ref{eq:53aa}) to be the basic
structure of the radiative correction, with the
``Appelquist-Carrazone" cancellation regarded as unnatural ``fine
tuning".  In this case it is the limit $M_i \rightarrow 0$ for
which $\delta v$ vanishes.  Without better control of the
quadratic-divergence issue, the situation appears to us to be
ambiguous.  We shall assume in what follows that the form of Eq.
(\ref{eq:53aa}) faithfully represents the true radiative
correction.

Either this radiative term, Eq. (\ref{eq:53aa}), is the
predominant contribution to the top of the ``Mexican hat'', or
else it is a correction. But we shall consider it to be
``unnatural" to suppose that this finite piece is cancelled off to
high accuracy by something else. Therefore one may expect that its
magnitude is limited by the known size of the Mexican-hat term.
This gives the rough bound
\begin{equation}
\frac{12}{64\pi^2} \sum^3_{i=1}\, H^2_i M^2_i \lsim (1 \ TeV)^2
\label{eq:53cc}
\end{equation}
leading to
\begin{equation}
7 \ TeV \gsim  \sqrt{\sum^3_{i=1} H^2_i M^2_i} \geq H_BM_B \geq
h_t M_B = 0.7 \, M_B \label{eq:53bd}
\end{equation}
where we have used our assumed lower bound on $H_B$ from Section
4.1.

The bottom line is that, together with the lower bound on $M_B$
coming from the precision electroweak measurements
\begin{equation}
|s_3| = \left|\frac{H_Bv}{M_B}\right| \lsim 0.1 \ , \label{eq:zg}
\end{equation}
leading to
\begin{equation}
M_B \geq 10\ H_Bv \geq 10\ h_tv = 10\ m_t = 1.7\ TeV \ ,
\label{eq:zh}
\end{equation}
we may infer
\begin{equation}
1.7 \ TeV \lsim M_B \lsim 10\ TeV \ .
\label{eq:zi}
\end{equation}
We may also infer that the remaining down-quarks are no heavier.
If the well-known symmetry relation connecting $m_b$ and $m_\tau$
\begin{equation}
\frac{m_\tau}{m_b} \approx \left[\frac{\alpha_s(m^2_{\rm
GUT})}{\alpha_s(m^2_b)}\right]^{0.5} \approx 0.3
 \label{eq:zj}
\end{equation}
can be applied to this heavy sector, then the generalization of
Eq. (\ref{eq:zi}) to the third-generation leptons $T$ would read
\begin{equation}
0.8\ GeV \lsim M_T \lsim 5\ TeV \ ,
\label{eq:zk}
\end{equation}
with the remaining leptons (other than possibly the gauge singlets
$N_i,N^\prime_i$) no heavier.

\section{Running Coupling Constants}

\subsection{The Fermion-Higgs Sector}

The presence of new heavy fermions, with assumed masses which are
not too heavy, implies that the usual considerations of the
scale-dependence of the dynamics needs to be modified.  In the
standard model, the top-quark Yukawa coupling is almost large
enough for it to attain strong coupling at an energy scale below
the GUT scale.  Also, the Higgs-boson quartic coupling will become
strong below the GUT scale unless its mass lies in the relatively
narrow window of 140--180  GeV.

It is of course a very central question whether there exists
``oases" in the desert, \ie\ energy scales small compared to the
GUT scale, where some subset of interactions become strong. Thus
far, the introduction of these new fermions has not demanded any
new oasis.  However, as we shall see, one or more oases can
 occur because of the extra loading of the
renormalization-group equations from the new degrees of freedom.

We first review the standard-model situation, working to one-loop
order, and neglecting small contributions, {\em e.g.} from gauge
degrees of freedom, whenever possible. The important quantities
are the top quark coupling $h_t \equiv h$, and the Higgs quartic
self-coupling $\lambda$. Ignoring all gauge couplings except the
important QCD correction, one has
\begin{equation}
\frac{dh^2}{dt} = 9h^4- 8g^2h^2
\label{eq:5a}
\end{equation}
and
\begin{equation}
\frac{d\lambda}{dt} = 12\, (\lambda^2+\lambda h^2-h^4)
\label{eq:5b}
\end{equation}
where our notation is
\begin{eqnarray}
t &=& \frac{1}{16\pi^2}\, \ell n\,
\frac{\mu^2}{\Lambda^2_{QCD}}\nonumber \\
 \frac{1}{g^2(t)} &=& \left(11 - \frac{2}{3}\, n_f\right) t \simeq
7t \\
\lambda(0.1) &=& \lambda_{EW} = \frac{m^2_H}{2v^2}\ ,\quad v
\approx 250\ GeV\, . \nonumber \label{eq:5c}
\end{eqnarray}
This gives for the scale parameter $t$ the values 0, 0.1, and 0.5
at the QCD, electroweak, and GUT scales respectively.

In our model, these equations become modified at scales large
compared to the scale $M$ of the heavy quark masses. The
modifications are as follows:
\begin{eqnarray}
\frac{dh^2}{dt} &=& 9h^4+6h^2 \sum^3_{j=1}\, H^2_j-8g^2h^2 \nonumber\\
\frac{dH^2_i}{dt} &=&
3H^4_i+6H^2_i\left(h^2+\sum^3_{j=1}H^2_j\right)
-8g^2H^2_i  \label{eq:5d} \\
\frac{d\lambda}{dt} &=& 12\lambda^2+12\lambda\,
\left(h^2+\sum^3_{j=1}H^2_j\right)-12
\left(h^4+\sum^3_{j=1}H^4_j\right)\ .\nonumber
\end{eqnarray}
We have not here included the contribution of the heavy leptons.
If their masses are related via an $SU(5)$-like symmetry to those
of the heavy quarks, then their Yukawa couplings $H_\ell$ can be
expected to be two to three times smaller than for their quark
partners (recall the $SU(5)$ expectation for the ratio of $\tau$
mass to $b$ mass) because of the QCD radiative correction
enhancing the quark mass. So the squared values will be 4 to 10
times smaller.  Finally there is a color factor 3 further favoring
the quarks over the leptons.  The bottom line is that it is
unlikely that inclusion of the leptons will change the behavior of
the remaining couplings significantly.

\begin{figure}[htbp]
\centering{\includegraphics{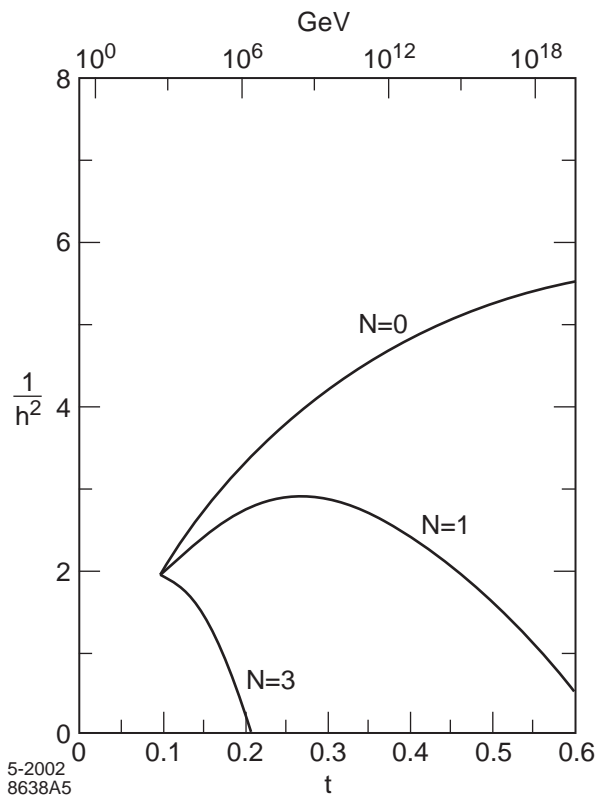}} \caption[*]{Running of the
top-quark Yukawa coupling for $N=0,1,$ and 3. \label{fig:5}}
\end{figure}

One sees in general from Eqs.  (\ref{eq:5d}) that the top quark
coupling $h^2$ will grow more rapidly than before, thanks to the
presence of the new terms.  And if the top quark coupling tends to
infinity, it will pull other couplings into the strong-coupling
regime, whenever those couplings are reasonably strong.  We shall
consider only the two extreme scenarios discussed in Section 4.1.
The flavor-symmetric scenario implies
\begin{equation}
H_1 = H_2 = H_3 \gsim h \ ,
\label{eq:zl}
\end{equation}
while the hierarchy scenario implies
\begin{equation}
H_1 \ll H_2 \ll H_3 \gsim h \ .
\label{eq:zm}
\end{equation}
We simplify matters further by assuming, for both scenarios,
\begin{equation}
H_3 = h \ .
\label{eq:zn}
\end{equation}
Then the first two expressions in Eq. (\ref{eq:5d}) reduce to the
single expression
\begin{equation}
\frac{dh^2}{dt} = (9+6N)h^4 - 8g^2h^2
\label{eq:5e}
\end{equation}
with $N=3$ for the flavor-symmetry option and $N=1$ for the
hierarchy option.  The running of $\lambda$ is then given by
\begin{equation}
\frac{d\lambda}{dt} = 12\left[\lambda^2+(N+1)h^2[\lambda
-h^2]\right] \ .\label{eq:5f}
\end{equation}
Equation (\ref{eq:5e}) can be integrated easily using the
expression for $g^2$ in Eq. (\ref{eq:5c}). The results are plotted
in Fig.~\ref{fig:5} for $N=0,1,$ and 3.

\vspace{.5cm}

We see that for the flavor-symmetry option, $N=3$, the heavy
quarks, top quarks, and Higgs bosons will be strongly coupled to
each other at an energy scale $\sim 10^3\ TeV$.  We therefore find
a new-physics oasis in the desert at this energy scale.  However,
for the hierarchy option, $N=1$, this does not appear to occur,
and the picture remains qualitatively similar to the minimal
standard model.

We again emphasize that the Higgs sector becomes strongly coupled
at a scale no larger than that for the quarks. If $h^2$ approaches
infinity, \ie\ into strong coupling, then the quartic coupling
$\lambda$ gets pulled with it. There is a special separatrix
solution, for which the quartic coupling at large values is in
proportion to $h^2$:
\begin{equation}
\lambda = K h^2 \quad (h^2 \gg g^2)
\label{eq:5g}
\end{equation}
with $K$ determined by the solution of
\begin{equation}
4K^2 + (1+2N)K-4(N+1) = 0 \ .
\label{eq:5h}
\end{equation}
This gives
\begin{equation}
K = \left\{ \begin{array}{rc}
0.88 & N=0 \\
1.09 & N=1\\
1.31 & N=3\end{array}\right\} = 1.1 \pm 0.2 \ .
\label{eq:5i}
\end{equation}

\begin{figure}[htbp]
\centering{\includegraphics[height=6.90in]{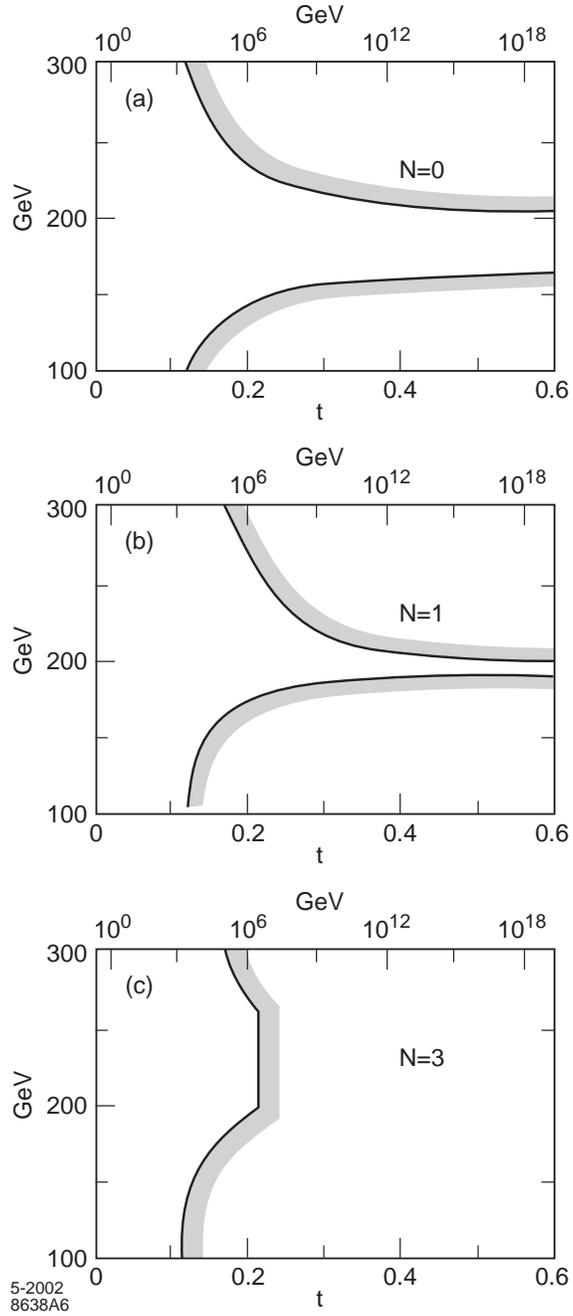}}
\caption[*]{Region of $m_H, t$ space where the theory remains
perturbative: (a) $N=0$ (standard model); (b) $N=1$; (c) $N=3$.
\label{fig:6}}
\end{figure}

If $\lambda$ is larger than this critical value, then the Higgs
sector can become strong at a scale less than where the top +
heavy-fermion sector gets strong.  If $\lambda$ is much smaller,
then it is driven through zero and to negative values, leading to
the unsatisfactory physics of metastable or unstable vacuum. This
is shown in Fig.~\ref{fig:6}. The $N=0$ case in
Fig.~\ref{fig:6}(a) represent the standard model, within the
approximations we have made.  Note that we obtain a value of $m_H
= 180 \pm 20\ GeV$ for the stability corridor, within which the
standard model remains perturbative up to the GUT scale.  This
value is higher by 20 GeV than the results of more sophisticated
analyses \cite{ref:B}.  The difference is our fault; we have made
several simplifications in the renormalization-group equations.

The hierarchical case, $N=1$, shown in Fig.~\ref{fig:6}(b), is
qualitatively similar.  The stability corridor is narrower, and
its center has moved upward by 5 GeV.  In the flavor-symmetric
case, $N=3$, shown in Fig.~\ref{fig:6}(c), the stability corridor
terminates at the oasis energy scale of 1000 $TeV$, where strong
coupling, or some other modification beyond the contents of this
model, is required.

In any case, the conclusion is that if at least some of the new
quarks have couplings to the Higgs boson equal to (or greater
than) that of the top quark, then their couplings, as well as the
Higgs self-coupling itself, get strong at an energy scale small
compared to the GUT scale. At higher energy scales the Higgs
degree of freedom need not exist at all, something perhaps welcome
from the point of view of the hierarchy problem and the
fine-tuning (or doublet triplet splitting) problem in dealing with
the Higgs as part of a GUT multiplet.  This feature makes this
``strong coupling" case an interesting hypothesis to investigate
in greater detail.

\subsection{Running of the gauge couplings}

The modifications to the renormalization-group equations for the
gauge bosons is straightforward.  We have
\begin{eqnarray}
 \frac{d}{dt}\left(\frac{1}{g_3}\right)^2 &=&
\left(11-\frac{2}{3}\cdot 6\right) = 7\qquad  \Rightarrow
\qquad \left(11-\frac{2}{3}\cdot
9\right) = 5 \nonumber \\[2ex]
 \frac{d}{dt} \left(\frac{1}{g_2}\right)^2 &=&
\frac{22}{3}-\frac{1}{3}\, \left[12+\frac{1}{2}\right] =
\frac{19}{6}\qquad \Rightarrow \qquad \frac{22}{3}-\frac{1}{3}\,
\left[18+\frac{1}{2}\right] = \frac{7}{6} \nonumber \\[2ex]
- \frac{d}{dt} \left(\frac{1}{g^\prime}\right)^2 &=& \frac{4}{3}\,
\left[5+\frac{1}{8}\right] = \frac{41}{6} \qquad \Rightarrow
\qquad \frac{4}{3}\, \left[\frac{15}{2}+ \frac{1}{8} \right] =
\frac{61}{6}\ .  \label{eq:52aa}
\end{eqnarray}
In the above equations, the first term in each of the square
brackets is the fermion contribution, and the second term is the
Higgs contributions. Our normalization of $g^\prime$ is such that
for $SU(5)$ unification
\[
\left( \frac{1}{g^\prime}\right)^2 \Rightarrow \frac{3}{5}\
\left(\frac{1}{g_1}\right)^2\ .
\]
\vspace{.5cm}
\begin{figure}[htbp]
\centering{\includegraphics{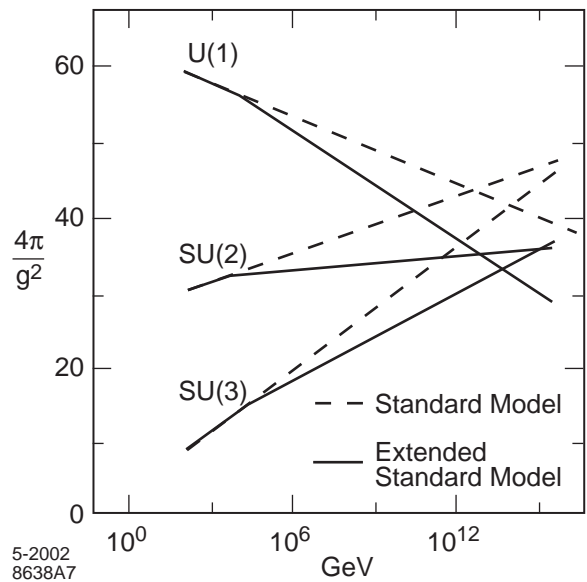}} \caption[*]{The running of
the gauge coupling constants. \label{fig:7}}
\end{figure}

With these changes we may compare the one-loop coupling-constant
evolution before and after the change. This is shown in Fig.
\ref{fig:7}.

We see that the three couplings converge somewhat better than in
standard $SU(5)$, but not as well as in the MSSM.

\section{Outlook}

The model we have described has been constructed in a relatively
unforced way. We began by assuming that the structure of the mass
and mixing matrices were simple and perturbative. This led
reasonably naturally to the hypothesis that this simplicity
reflected a simple mixing pattern occurring only in the down-quark
sector, one that led to a suggestive albeit not compelling
right-angle structure of the unitarity triangle. This in turn led
to introduction of heavy down quarks in order to implement the
mixing. The constraint of anomaly cancellation then led to the
extension to $E(6)$ multiplets for the elementary fermions.
Examination of the Higgs sector led to the feature of radiatively
induced Mexican-hat structure of the Higgs potential via the new
heavy down quarks. In addition, the phenomenology has the feature
that  flavor and CP violation are infrared, diminishing rapidly
with higher energy until an energy scale is reached where the mass
terms involving the heavy quark sector become dynamical.

All of this we find conceptually attractive. And the story may not
end at this point. There is an especially interesting extension
associated with the hierarchical choice of new Higgs couplings, as
discussed in the previous section. If that choice is made, there
exists a parity symmetry within the Higgs sector, broken only by
the parity violating electroweak gauge symmetry. To see this we
need only invoke the $SU(3) \times SU(3) \times SU(3)$ breakdown
of $E(6)$ \cite{rosner}, where the quark nonets are
\begin{equation}
q_L = \left(\begin{array}{c} u\\ d\\ D\end{array}\right)_L \qquad
q_R = \left(\begin{array}{c} u\\ D\\ d\end{array}\right)_R \
.\label{eq:xa}
\end{equation}
In this notation, Eqs. (\ref{eq:3f})  and (\ref{eq:3k}) become
\begin{equation}
\L^\prime = \sum^3_{i=1} \bar q^i_L \Phi \left(\begin{array}{cc}
h_i & 0 \\ 0 & H_i\end{array}\right)\, q^i_R + \mbox{h.c.}
\label{eq:xb}
\end{equation}
where the Higgs field and explicitly shown matrix both act in the
upper $2\times 2$ block. With the hierarchical assumption
\begin{equation}
h_i = H_i \  \label{eq:xc}
\end{equation}
we see that the Higgs part of the action is parity invariant in
the absence of the mass terms. So, ignoring the corrections coming
from the parity-violating electroweak gauge symmetry, parity
conservation becomes restored at energies large compared to the
mass scale of the heavy down-quarks. Note also that all the
mass-mixing terms are encapsulated neatly within a ($3,\bar 3$)
Higgs representation:
\begin{equation}
M_{ij} = \left( \begin{array}{ccc} h_iv & 0 & 0 \\
0 & h_iv & M_i \\ 0 & M_i & \mu_{ij} \end{array}\right) \
.\label{eq:xd}
\end{equation}
The identification of the new heavy-Higgs couplings with the
light-Higgs couplings leads to additional constraints on masses
and mixings. For example, from Eqs. (\ref{eq:3i}) and
(\ref{eq:3k}), we have
\begin{equation}
m_{ij} = \frac{H_iv}{M_i}\, \mu_{ij} = h_iv
\left(\frac{\mu_{ij}}{M_i}\right) \ . \label{eq:xe}
\end{equation}
Therefore
\begin{eqnarray}
\frac{m_d}{m_u} &=& \frac{\mu_{11}}{M_1} = 1.8 \nonumber \\[.5ex]
\frac{m_s}{m_c} &=& \frac{\mu_{22}}{M_2} = 0.12 \\[.5ex]
\frac{m_b}{m_t} &=& \frac{\mu_{33}}{M_3} = 0.03 \nonumber \ .
 \label{eq:xf}
\end{eqnarray}
We are {\em assuming} that the off-diagonal flavor mixing effects
are perturbative, so that the diagonal elements of the mass matrix
are in fact the measured masses. The off-diagonal elements are
known in terms of diagonal elements:
\begin{equation}
\frac{\mu_{ij}}{\mu_{ii}} = \frac{m_{ij}}{m_{ii}} \ .
\label{eq:xg}
\end{equation}
From the estimates in Eq. (\ref{eq:2m}), we obtain
\begin{eqnarray}
\left|\frac{\mu_{12}}{\mu_{11}}\right| &=& \frac{35\
\mbox{MeV}}{7\ \mbox{MeV}} \approx 5 \nonumber\\[.5ex]
\left|\frac{\mu_{13}}{\mu_{11}}\right| &=& \frac{15\
\mbox{MeV}}{7\ \mbox{MeV}} \approx 2 \\[.5ex]
\left|\frac{\mu_{23}}{\mu_{22}}\right| &=& \frac{m_{23}}{m_{22}}
\approx 1.2 \ . \nonumber \label{eq:xh}
\end{eqnarray}
Actually these relations are completely general.

Now we have demanded that (Eq. (\ref{eq:zi}))
\begin{equation}
M_3 \lsim 10\ \mbox{TeV} \label{eq:xi}
\end{equation}
so that
\begin{equation}
\mu_{33} = 0.03\, M_3 \leq 300\ \mbox{GeV} \ . \label{eq:xj}
\end{equation}
But (cf. Eq. (103))
\begin{equation}
\mu_{11} \approx 1.8\, M_1 \geq (1.8)(130\, \mbox{GeV}) = 240\,
\mbox{GeV} \label{eq:xk}
\end{equation}
implying that the three diagonal $\mu_{ii}$ are almost the same.
This suggests
\begin{equation}
M_2 = \frac{\mu_{22}}{0.12} \sim 2\, \mbox{TeV} \ . \label{eq:xl}
\end{equation}

Assuming here that the $\mu_{ii}$ are exactly the same
\begin{equation}
\mu_{ii} \cong 300\, \mbox{GeV} \equiv \mu \qquad\qquad i = 1,2,3
\label{eq:xm}
\end{equation}
leads to a mass matrix with corrections which go beyond our
perturbation theory assumptions. To see this, write
\begin{equation}
M^\prime_{ij} = \frac{m_i}{M_i}\, \mu_{ij} = \frac{m_i\mu}{M_i}\,
u_{ij} \label{eq:xn}
\end{equation}
with
\begin{equation}
u_{ij} = \frac{\mu_{ij}}{\mu_{ii}} = \frac{\mu_{ij}}{\mu} \ .
\label{eq:xo}
\end{equation}
Then the square of $M'$ is
\begin{equation}
\M^{\prime^{2}}_{ij} = \left(M^\prime
M^{{\prime}^\dagger}\right)_{ij} = m_{ii} m_{jj}
u_{ik}u^*_{jk} \label{eq:xp}
\end{equation}
and we see that
\begin{equation}
(\M^{\prime^{2}})_{ii} = (m_{ii})^2 \left[1+\sum_{k\ne i}
|u_{ik}|^2\right]
\gg m^2_{ii} \ . \label{eq:xq}
\end{equation}
Therefore we need to start from scratch and study the
diagonalization with much greater care. This does not imply an
unworkable scenario, only one which goes in a somewhat different
direction than what we have heretofore set up. Going further,
however, is left for future work.

\section{Conclusions}

The main features for experiment of this model are as follows:

\begin{enumerate}

\item
The possibility of ``Stech texture" for the mass matrix, leading
to an approximate right angle ($\gamma$) in the unitarity
triangle.

\item
The existence of three generations of heavy electroweak-singlet
down- quarks which decay into their light counterparts plus $W$,
$Z$, or Higgs. The masses should be no larger than roughly 10 TeV.
The leptons most reasonably are a factor two or so lighter than
their heavy-quark counterparts. The first generation quark masses
may be near the experimental bound, which is 130 GeV.

\item
Flavor and CP violation are induced only by mass-mixing. Therefore
above that mass scale, such effects rapidly diminish, only
re-emergent if and when the mechanism for the relevant mass terms
becomes dynamical. This also implies that radiative-correction
effects are in all cases not divergent.

\item
Some precision electroweak observables are in principle sensitive
to the existence of these new degrees of freedom. The ordinary
down quarks and leptons are mixed slightly with their heavy
counterparts, making them to not transform as pure doublets or
singlets. This leads to nonuniversality of the asymmetries
measured in electron-positron annihilation processes. On the other
hand, a variety of one-loop radiative correction effects in the
down-quark or lepton sectors vanish. No significant corrections,
for example, are expected in mass mixing of kaons or neutral
$B$'s, nor in $\epsilon$ and $\epsilon'$, nor in $b \rightarrow s
\gamma, \mu \rightarrow e \gamma$, and $K_L \rightarrow \mu e$,
nor in the unitarity triangle. There can be significant effects in
the up-quark sector, e.g. in $D^0-\overline{D^0}$ mixing.

\item
If the Higgs couplings to the heavy quarks are flavor universal,
and at least as large as the Higgs coupling to the top quark, then
there will be an oasis in the desert, at an energy scale of about
1000 TeV, where the Higgs, top-quark, and heavy down-quark
couplings all become strong. Additional new physics is then
assured above this energy scale.

\item
If the Higgs couplings to the new quarks are hierarchical, then
there need be no oasis. If it is postulated that none exists, the
Higgs boson must have a mass of 160 $\pm$ 20 GeV.

\end{enumerate}

\section*{{\bf Acknowledgments}}

We would like to thank J. Hewett and T. Rizzo for extensive help
on many aspects and  for many useful discussions. We also thank M.
Peskin and X. Tata  for useful discussions. This work was
supported in part by the U. S. Department of Energy under contract
number DE--AC03--76SF00515 and grant number DE-FG-03-94ER40833.


\begin{thebibliography}{99}
\bibitem{ref:x}
This is obtained from an  analysis of $b^\prime$ decaying into
jets by the D\O\  Collaboration; see e.g. Ref. [21].
\bibitem{hewett1}
J. Hewett, T.G. Rizzo and J.A. Robinson, {\em Phys. Rev.} {\bf
D33}, 1476 (1986).  It is shown here that if $E(6)$ is broken in
the pattern $E(6) \rightarrow SO (10) \times U(1) \rightarrow
SU(3) \times SU(2)  \times U(1)  \times U(1) \rightarrow SU(3)
\times SU(2)  \times U(1)$ then $X$ has the value + 18.
\end{thebibliography}
\end{document}